\documentclass[aps,prl,twocolumn,superscriptaddress,showpacs,amsmath,floatfix,citeautoscript]{revtex4-2}

\usepackage{amssymb}
\usepackage{amsmath}
\usepackage{colordvi}
\usepackage[colorlinks]{hyperref}
\usepackage{amsthm}
\usepackage{subeqnarray}
\usepackage{cases}
\usepackage{enumerate}
\usepackage{graphicx}
\usepackage{epsfig}
\usepackage{epstopdf}
\usepackage{subfigure}
\usepackage{color}
\usepackage{ulem}

\def\avg(#1){\langle#1\rangle}

\def\be{\begin{equation}}
\def\ee{\end{equation}}
\def\bea{\begin{eqnarray}}
\def\eea{\end{eqnarray}}

\begin{document}
\title{A quantum spin liquid phase in the Kitaev-Hubbard model}

\author{Shaojun Dong}
\affiliation{Institute of Artificial Intelligence, Hefei Comprehensive National Science Center}

\author{Hao Zhang}
\affiliation{CAS Key Laboratory of Quantum Information, University of Science and Technology of China, Hefei 230026, People's Republic of China}
\affiliation{Synergetic Innovation Center of Quantum Information and Quantum Physics, University of Science and Technology of China, Hefei 230026, China}

\author{Chao Wang}
\affiliation{Institute of Artificial Intelligence, Hefei Comprehensive National Science Center}

\author{Meng Zhang}
\affiliation{CAS Key Laboratory of Quantum Information, University of Science and Technology of China, Hefei 230026, People's Republic of China}
\affiliation{Synergetic Innovation Center of Quantum Information and Quantum Physics, University of Science and Technology of China, Hefei 230026, China}

\author{Yong-Jian Han}
\email{smhan@ustc.edu.cn}
\affiliation{CAS Key Laboratory of Quantum Information, University of Science and Technology of China, Hefei 230026, People's Republic of China}
\affiliation{Institute of Artificial Intelligence, Hefei Comprehensive National Science Center}
\affiliation{Synergetic Innovation Center of Quantum Information and Quantum Physics, University of Science and Technology of China, Hefei 230026, China}

\author{Lixin He}
\email{helx@ustc.edu.cn}
\affiliation{CAS Key Laboratory of Quantum Information, University of Science and Technology of China, Hefei 230026, People's Republic of China}
\affiliation{Synergetic Innovation Center of Quantum Information and Quantum Physics, University of Science and Technology of China, Hefei 230026, China}
\affiliation{Hefei National Laboratory, University of Science and Technology of China, Hefei 230088, China}

\begin{abstract}
The quantum spin liquid (QSL) state has been searched intensively in Kitaev-like materials, such as the Iridium oxides $A_2$IrO$_3$ and $\alpha$-RuCl$_3$.
The half-filled Kitaev-Hubbard model with bond dependent hopping terms is used to describe the Kitaev-like materials, which is calculated using the state-of-the-art fermionic projected entangled pair states (fPEPS) method.
We find a QSL phase near the Mott insulator transition, which has a strong first-order transition to the semi-metal phase with the decrease of Hubbard $U$.
We suggest that a promising routine to find the QSL  is to find the  Iridium oxides that are near the Mott insulator transitions.
\end{abstract}
\maketitle


A quantum spin liquid (QSL)~\cite{Anderson1973,Anderson1987,Balents2010} state is a quantum state that lacks any long range magnetic order even down to zero temperature.
QSLs have nontrivial topological properties
that may host exotic excitations with fractional statistics, such as spinons, and visions, etc.,
which may have important applications in quantum computing\cite{Kitaev2003,Kitaev2006} and may play a crucial role in high-temperature superconductivity.

The Kitaev model\cite{Kitaev2006} is an exactly solvable model on a 2D honeycomb
lattice, which hosts a QSL ground state. Several Iridium oxides $A_2$IrO$_3$, as well as $\alpha$-RuCl$_3$, have
been proposed to realize the Kitaev QSL~\cite{Choi2012,Singh2012,Plumb2014,Sears2015,Baek2017,Wolter2017,Sears2017,Gass2020,Bachus2020}.
These materials have a honeycomb structure and the strong spin–orbit coupling leads to an effective $j_{\rm eff}$=$\frac{1}{2}$ spin model
with bond dependent anisotropic exchange interactions\cite{Plumb2014,Agrestini2017}, which are the essential ingredients of the  Kitaev model.
In addition to the Kitaev exchange interactions, there are also  Heisenberg interactions in these materials~\cite{Chaloupka2010}. The Kitaev-Heisenberg model has been intensively studied, and it has been shown that the QSL can only survive in a rather
small parameter space~\cite{Chaloupka2013,Chaloupka2010,Schaffer2012,Reuther2011,Sela2014}. Indeed, Na$_2$IrO$_3$ and $\alpha$-RuCl$_3$ were found to have a zigzag antiferromagnetic (AFM) order by resonant X-ray magnetic scattering and inelastic neutron scattering experiments.
Tremendous efforts have been made to find the QSL in these materials, and yet no evidence of QSL has been found so far\cite{Liu2011,Ye2012,Choi2012,Sears2015}.
An important question is that given the extremely small parameter space for the QSL in the Kitaev-Heisenberg model, is it even possible
to find the Kitaev QSL in real materials?

The Kitaev-Hubbard model is a more realistic model to describe the Iridium oxides.
When the Hubbard $U$ is small, higher order interactions become important, which may introduce
exotic states.  The Kitaev-Hubbard model has been studied by mean-field theories~\cite{Hassan2013,Liang2014,Faye2014}.
It has been shown that there exists a QSL phase in the region of  $t' <t$ when $U$ is small,
where $t$ and $t'$ are the isotropic and  spin-dependent hopping terms respectively. A further decrease in $U$ results in a semi-metal (SM) phase.
However, these calculations were based on mean-field approximations~\cite{Hassan2013,Liang2014,Faye2014},
which need to be examined by more rigorous methods.
Furthermore, these studies focus on the  $t'<t$ region, and the phase diagram for $t'>t$ was missing.
Experimentally, the Iridium oxides materials, e.g., Na$_2$IrO$_3$ and $\alpha$-RuCl$_3$, are believed to have strong spin-dependent hopping terms\cite{Sears2015,Banerjee2016,Shubhajyoti2019,Rau2014}, and in the $t'>t$ region.

The projected entangled pair states method (PEPS)~\cite{Lubasch2014,Roman2014IntroTNS,Verstraete2008,Xiang2008,Verstraete2004,Liu2017}, and its generalization to fermionic systems (fPEPS)~\cite{Dong2019fTensor,Gu2010,Corboz2010,Kraus2010,Barthel2009} provide systematically improvable variational wave functions for the many-body problems, which allow more rigorous treatment of the Kitaev-Hubbard model.
In this Letter, we apply this recently developed and highly accurate fPEPS method to explore the phase diagram of the half-filled Kitaev-Hubbard model.
The results show that the QSL state is absent in the $t'<t$ region in contrast to previous mean-field results~\cite{Hassan2013,Liang2014,Faye2014}. Instead, we find a QSL phase in the $t'>t$ region when $U$ is small.
We show that the phase transition from the SM phase to the QSL phase is a first-order transition, whereas the QSL-zigzag transition is a continuous transition.
Given that the Iridium oxide materials and $\alpha$-RuCl$_3$ are in the $t'>t$ region, it is possible to find suitable materials that may host the QSL.

\begin{figure} [tbp]
		\begin{center}
		\includegraphics[width=0.48\textwidth]{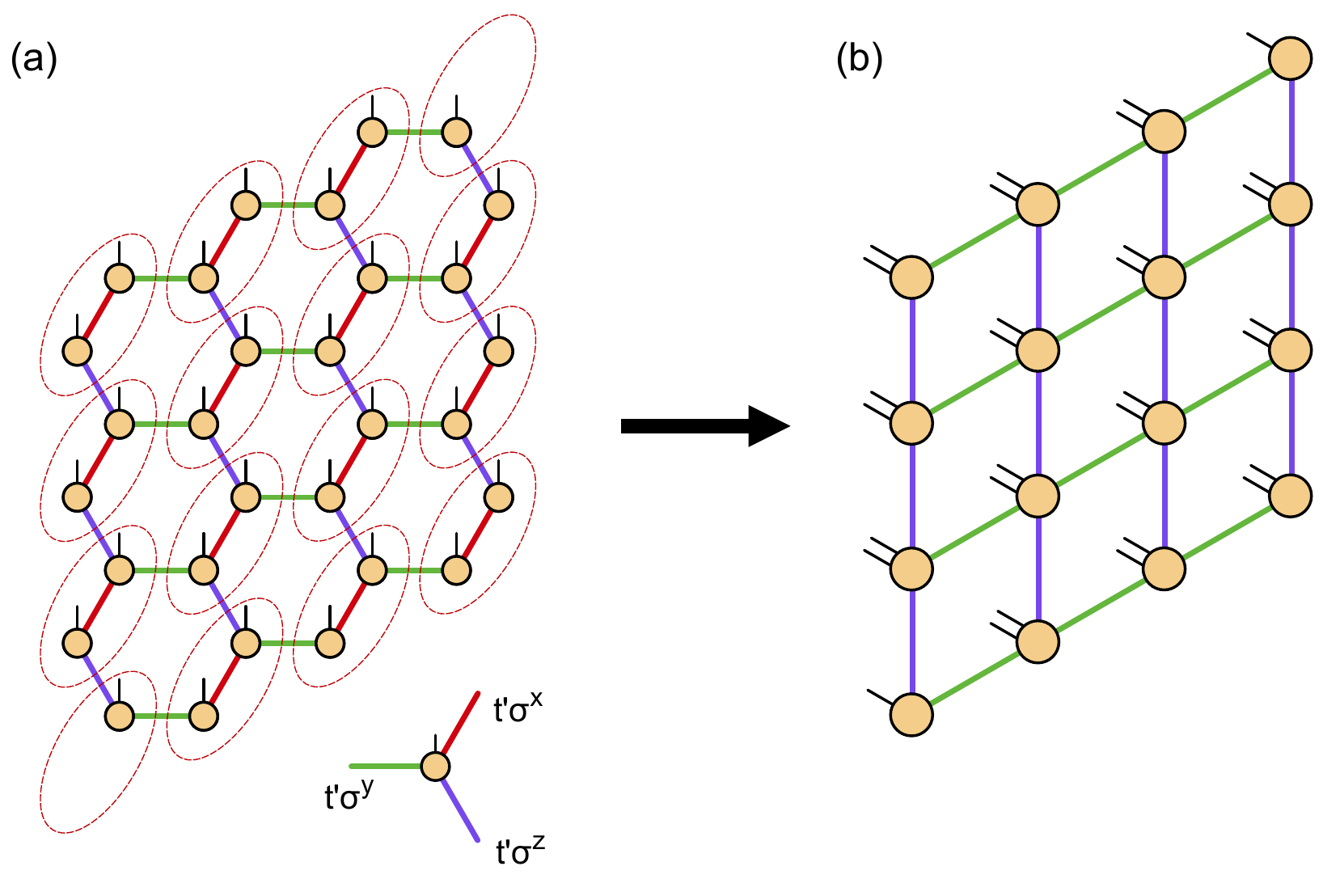}
		\caption{(Color online) (a) The honeycomb lattice with bond dependent interactions.  (b) The honeycomb lattice is mapped to a
PEPS defined on a square lattice, where two physical indices on the honeycomb lattice are combined in a single site.
}
\label{fig:TNS}
		\end{center}
\end{figure}

The Hubbad-Kitaev model reads,
\begin{equation}\label{eq:Hubbad-Kitaev}
H=\sum_{\langle i ,j \rangle_\alpha,s} \{ \hat{c}_{i,s}^\dag (\frac{t+t'\sigma^\alpha}{2})\hat{c}_{j,s} + H.c. \} + U\sum_i \hat{n}_{i\uparrow} \hat{n}_{i\downarrow} ,
\end{equation}
where $\hat{c}_{i,s}$ is the annihilation operator that destroys an electron with spin $s$ at site $i$, and $\hat{n}_{i,s}=\hat{c}_{i,s}^\dag \hat{c}_{i,s}$ is the number operator.
$\sigma^\alpha$, with $\alpha$=$x$, $y$, $z$ are the pauli matrices. $\langle i,j \rangle_\alpha$ denotes the nearest-neighbor pairs in the three hopping directions of the lattice, as sketched in Fig.~\ref{fig:TNS}. The $t$ and $U$ terms are the hopping and the on-site Coulomb  interaction terms in the normal Hubbard model, whereas the $t'$ is the spin-dependent hopping due to spin-orbit coupling.

The fPEPS method~\cite{Lubasch2014,Roman2014IntroTNS,Verstraete2008,Xiang2008,Verstraete2004,Liu2017,Dong2019fTensor} is one of the  most promising methods to study strongly correlated electron systems.
We simulate the honeycomb lattice with open boundary condition (OBC) using a square tensor network. We map the honeycomb lattice from Fig.~\ref{fig:TNS}(a), to a square lattice of Fig.~\ref{fig:TNS}(b), where two sites of the honeycomb lattice  are treated as a single site in the square lattice (except those sites at the corner, which remain a single site in the square lattice). Each site of tensor network contains two physical indices. The physical properties can still be easily calculated on the physical (honeycomb) lattice. We optimize the wave functions using the so called fPEPS++ method developed in our group~\cite{Liu2017,Dong2019fTensor,Liu2018}, i.e., the fPEPS wave functions are optimized via a stochastic gradient method,
whereas the energy and energy gradients are calculated using a  Monte Carlo sampling method.
 This method significantly reduces the computational complexity with respect to the bond dimension $D$, thereby allowing a much larger bond dimension to be used, resulting in highly accurate and converged results for large finite systems.

  All the parameters in the fPEPS wave functions are independent and subject to optimization.  We first obtain the quantum wave functions with
  the simple update method, and the fPEPS wave functions are further optimized using  the stochastic gradient method until the results fully converge~\cite{Liu2017,Dong2019fTensor,Liu2018}.
  In our calculations,  the bond dimension $D$=14 and the truncation bond dimension $D_c$=42 are used, which show good convergence.

The magnetic phase is determined by the spin structure factor,
\begin{equation}\label{AFM_factor}
S(\{\boldsymbol{k}_\alpha\})=\sum_{i,j,\alpha} e^{-i\boldsymbol{k_\alpha} \cdot (\boldsymbol{r_i}-\boldsymbol{r_j})}S^\alpha_i \cdot S^\alpha_j ,
\end{equation}
where $\boldsymbol{r_i}$ is the coordinate of the honeycomb lattice and $S^\alpha_i$ is the spin operator at site $\boldsymbol{r_i}$, with $\alpha=x,y,z$. The  AFM
order is characterized by the non-zero value of the spin structure factor at $\boldsymbol{k}_x=\boldsymbol{k}_y=\boldsymbol{k}_z=(\frac{4\pi}{3},0)$, whereas the zigzag order is detected by the spin structure at  $\boldsymbol{k_x}=(\frac{2\pi}{3},0)$, $\boldsymbol{k_y}=(-\frac{\pi}{3},\frac{\sqrt{3}\pi}{3})$ and $\boldsymbol{k_z}=(\frac{\pi}{3},\frac{\sqrt{3}\pi}{3})$.  The QSL states are distinguished when all these spin orders vanish but
still have finite charge gaps,
\begin{equation}\label{gap}
\Delta=E_{N+1}+E_{N-1}-2E_N,
\end{equation}
where $E_N$ is the total energy of the system with $N$ electrons.



We first discuss the Kitaev-Hubbard model Eq. (1) in the large-$U$ limit.
Without loss of generality, we take $t=1$ throughout the paper.
At half-filling, the model can be reduced to the Kitaev-Heisenberg spin model
to the leading order of $1/U$~\cite{Hassan2013,Faye2014},
\begin{equation}
H_{\rm eff}=\sum_{\langle i ,j \rangle_\alpha} (\frac{(1-t'^2)}{U} {\bf S}_i  \cdot  {\bf S}_j + \frac{2t'^2}{U}S_i^\alpha S_j^\alpha) \,,
\label{eq:heff}
\end{equation}
where $S_i^\alpha,\alpha=x,y,z$ are the spin operators at site $i$ the ${\bf S}_i=(S_i^x,S_i^y,S_i^z)$ , and the $\langle i,j \rangle_\alpha$ denotes the nearest-neighbor pairs in the three hopping directions of the lattice (see Fig.~\ref{fig:TNS}).
The Kitaev-Heisenberg model
has been studied intensively, and the phase diagram of the model is well known~\cite{Chaloupka2010,Reuther2011,Schaffer2012,Okamoto2013,Steinigeweg2016,Janssen2016,Gohlke2017,
Joshi2018,Metavitsiadis2019,Consoli2020,Morita2020,Zhang2021}.
In Ref.~\onlinecite{Chaloupka2013},
the authors extended the original model to its full parameter space, i.e.,
\begin{equation}
H_{\rm KH}=\sum_{\langle i ,j \rangle_\alpha} [J {\bf S}_i \cdot {\bf S}_j + K S_i^\alpha S_j^\alpha ]\, ,
\label{eq:hkh}
\end{equation}
where $J$=$\cos\phi$ is the Heisenberg coupling strength, and  $K$=2$\sin\phi$ is the Kitaev coupling strength.
The phase angle $\phi$ may vary from 0 to $2\pi$.
The phase boundaries have been obtained by exact diagonization of the Hamiltonian on a 24-site hexagonal lattice
with periodic boundary conditions.  Compared with the  large-$U $ effective Hamiltonian of Eq.~\ref{eq:heff}, the angle $\phi$ can be related to the $t'$ as $\cot \phi=\frac{1-t'^2}{t'^2}>-1 $ and $\sin\phi=\frac{t'^2}{U}>0$, and therefore we have a constrain of $0\le \phi <3\pi/4$ for Eq. \ref{eq:hkh}. In this parameter region, there does exit a QSL phase on $\phi\in (88^o,92^o)$, corresponding to $t' \sim (0.99970,1)$ in Eq. ~\ref{eq:heff}, which is almost a single point in the parameter space.
It has been shown that when $t'<1$, the ground state is an AFM phase and is the zigzag  phase for $t'>1$.
Therefore, in the large $U$ limit, the AFM phase and the zigzag phase are separated by a QSL state that survives (almost) only at the $t'=1$ line.

The decrease in $U$ may introduce higher order spin interactions~\cite{Hassan2013}, which may stabilize the QSL phase in a larger region.
Several studies have shown some insight into this problem, where the authors claim that an algebraic QSL lies between $t'\sim 0.7$ and $t'=1$, when $U $ is small~\cite{Hassan2013,Liang2014,Faye2014}.
However, these calculations were based on mean-field approximations~\cite{Hassan2013,Liang2014,Faye2014},
which need to be examined by more rigorous method.
Furthermore, the studies focused on the  $t'<$1 region, and the phase diagram for $t'>1$ was not studied.

\begin{figure} [tbp]
		\begin{center}
		\includegraphics[width=0.48\textwidth]{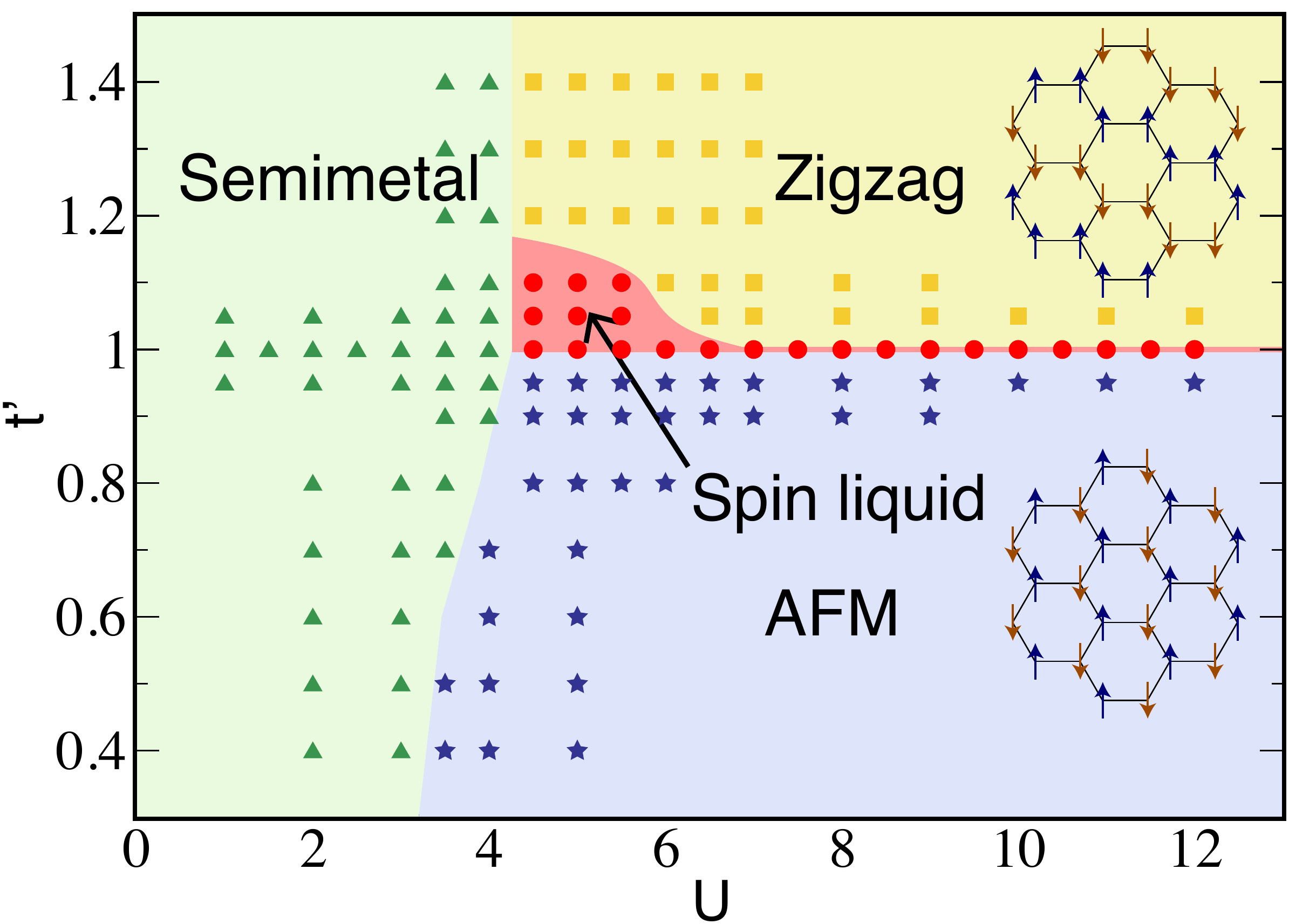}
		\caption{(Color online) The ground state phase diagram of the Kitaev-Hubbard model on the $t'$-$U$ plane, where we have set $t$=1.  Four phases have been identified including a SM phase, an AFM phase,  a zigzag phase and a QSL phase.  The scatters represent the parameters that are calculated using the fPEPS method.
}
\label{fig:PhaseDiagram}
		\end{center}
\end{figure}

\begin{figure} [tbp]
		\begin{center}
		\includegraphics[width=0.45\textwidth]{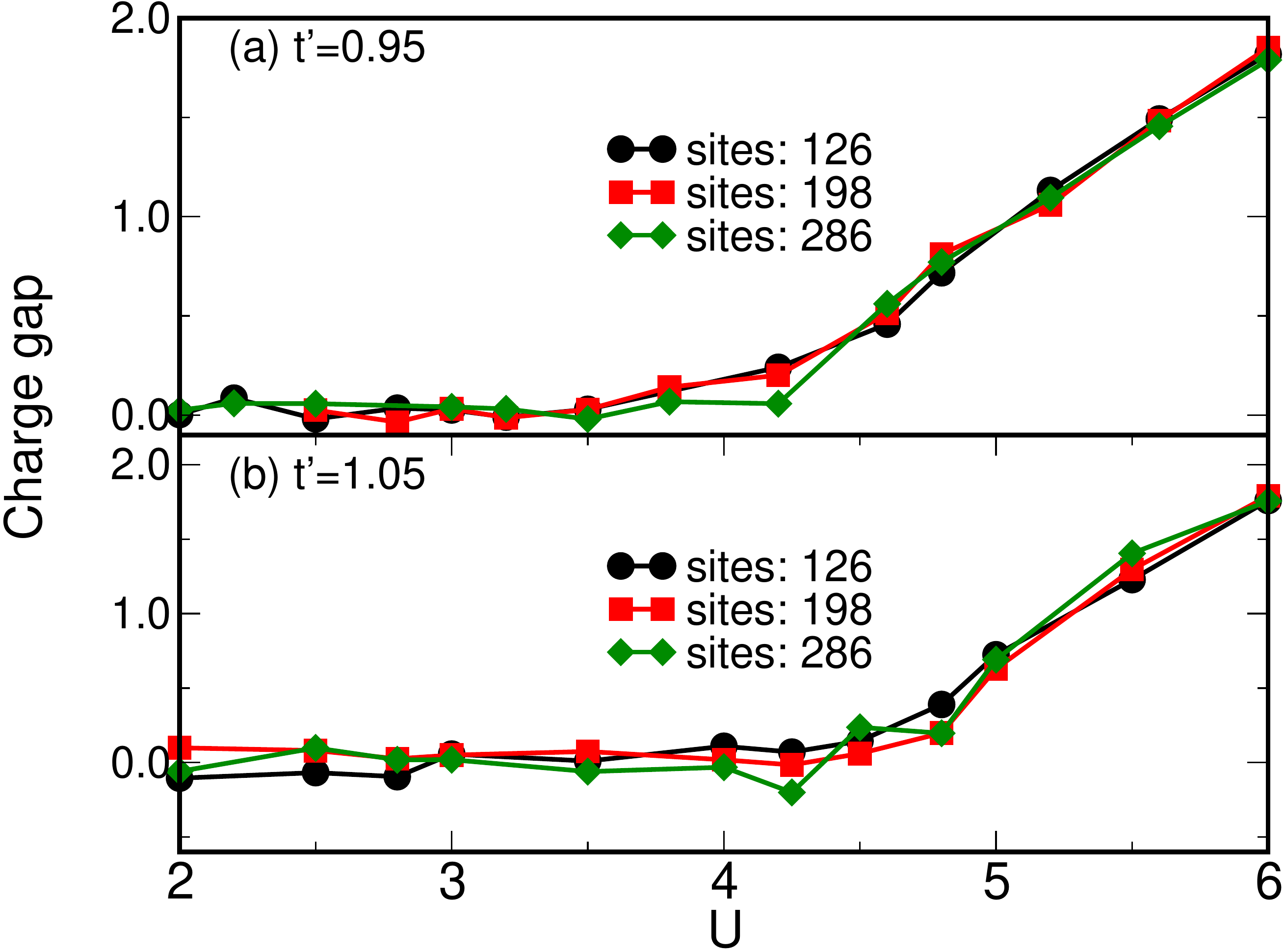}
\		\caption{  (Color online) The charge gaps $\Delta$ on different lattice sizes for (a) $t'=0.95$ and (b) $t'=1.05$. }
\label{fig:t095}
		\end{center}
\end{figure}

We calculate the phase diagram of the Kitaev-Hubbard model in the $t'$-U plane,
using the fPEPS method, and the results are shown in Fig.~\ref{fig:PhaseDiagram}.
Four phases have been identified in the phase diagram.
On the left side of the diagram, where $U$ is small, there is a large SM phase that adiabatically connects to the phase at $t'$=0, and $U$=0, i.e.,
the electronic structure of graphene.
In the large $U$ limit, the system is in an AFM phase for $t'<$1, and a zigzag phase, when $t'>$1.
Remarkably, there is a QSL phase in the parameter range $t'\in (1,1.2)$ and $U\in (4.5,7)$.
For $U>7$, the QSL phase reduces into the $t'$=1 line, consistent with the results of the Kitaev-Heisenberg model\cite{Chaloupka2013}.

The SM phase is accompanied by the vanishing of the charge gap.
Figure~\ref{fig:t095}(a), (b) depict the charge gaps at $t'$=0.95 and 1.05, on the honeycomb lattices of 126, 198, 286 sites. The charge gaps decrease with decreasing $U$. For both $t'$=0.95 and $t'$=1.05, the charge gaps become zero at
approximately $U \sim$ 4.5, which are the phase boundaries between the SM and Mott insulator phases.
The SM phase is further ensured with the vanishing of the local magnetic moment.

\begin{figure*} [tbp]
		\begin{center}
		\includegraphics[width=0.8\textwidth]{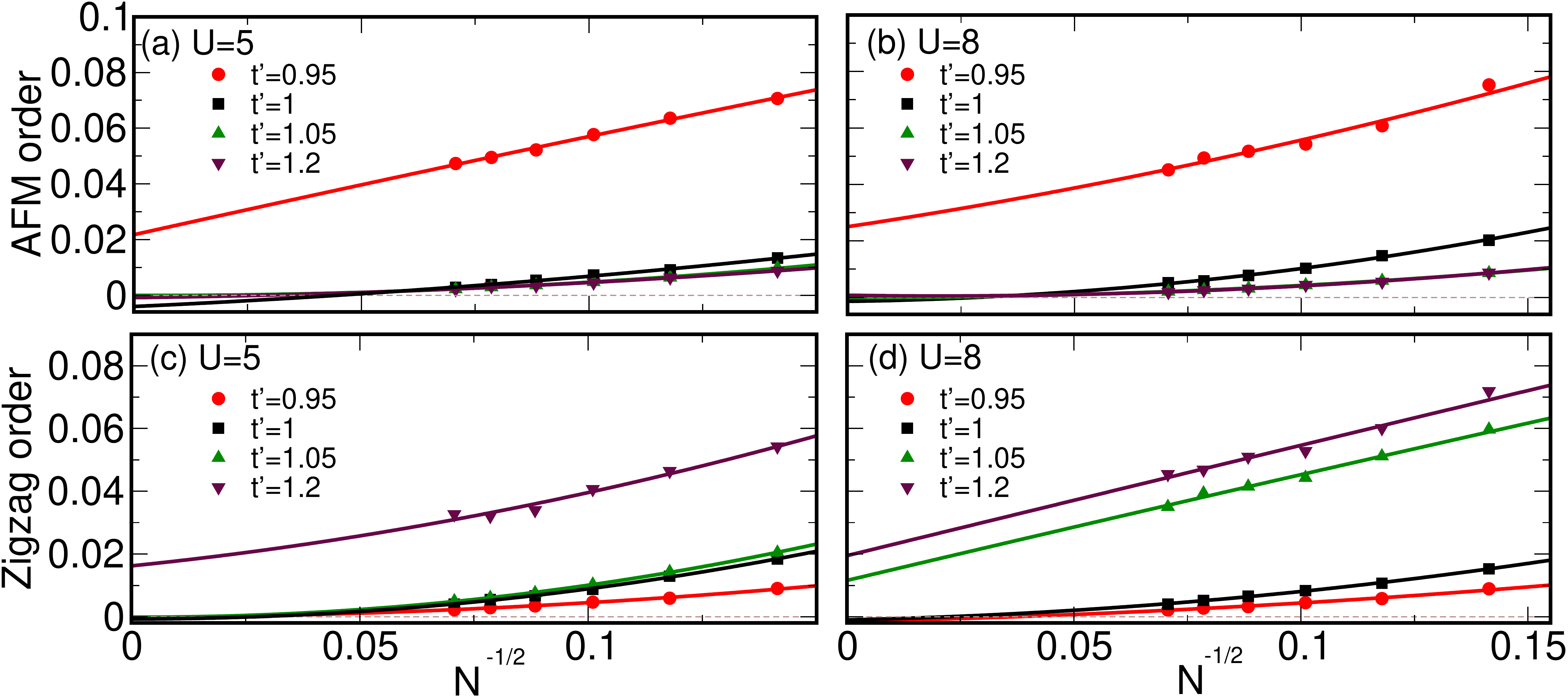}
\		\caption{(Color online) The finite size scaling of the AFM order parameters for (a) $U$=5 and (b) $U$=8. The  finite size scaling of the zigzag order parameters
is shown in (c) for $U$=5 and (d) for $U$=8.
}
\label{fig:extrapolate}
		\end{center}
\end{figure*}

We now focus on the insulating region. To determine the magnetic order, we calculate the AFM
and zigzag order parameters in the thermodynamic limit via finite size scaling.
Figure \ref{fig:extrapolate}(a), (c) depict AFM and zigzag order parameters for $U$=5, with $t'$=0.95, 1, 1.05 and 1.2 as functions of
the square root of the number of lattice sites used in the calculations,
whereas Fig.~\ref{fig:extrapolate}(b), (d) show  the results for $U$=8.
To reduce the boundary effects, the order parameters are calculated using only the central region of the lattice\cite{Stoudenmire2012,Liu2017,Liu2018}.
For $U$=5 and $t'$=0.95, the system shows a finite AFM order in the thermodynamic limit, and when $t'$=1.2, the system shows a zigzag order.
However, for  $t'$=1 and $t'$=1.05, both AFM and zigzag orders vanish in the  thermodynamic limit.
In contrast, for $U$=8 and  $t'$=0.95, the system also has an AFM order, and for $t'$=1.05 and $t'=1.2$, the system shows a zigzag order. Only when $t'$=1, do both AFM and zigzag orders vanish, as expected from the Kitaev-Heisenberg model.\cite{Chaloupka2013}

\begin{figure} [tbp]
		\begin{center}
		\includegraphics[width=0.45\textwidth]{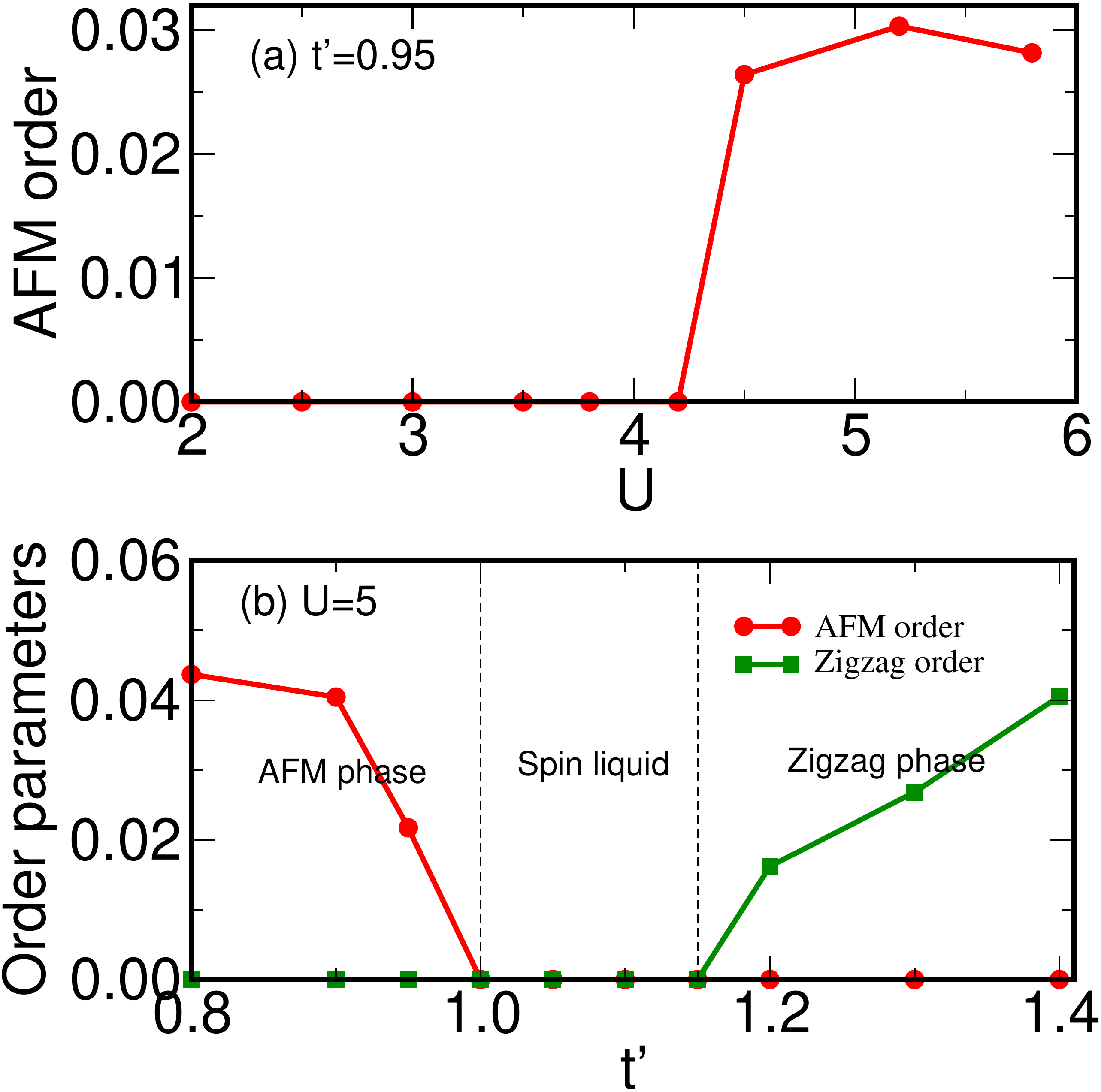}
		\caption{(Color online) (a) AFM order in the thermodynamics limit as a function of $U$ for $t'$=0.95.
(b) AFM order and zigzag order as functions of $t'$ at $U$=5.
}
\label{fig:SpinOrder}
		\end{center}
\end{figure}

Figure~\ref{fig:SpinOrder}(a) depicts the AFM order parameter (in the thermodynamic limit) as a function of $U$ for $t'$=0.95.
The AFM order disappears at approximately $U \sim$4.5, which is coincident with the disappearance of the charge gap $\Delta$.
This result suggests that there is no other phase between the SM and AFM phases.
The calculated phase boundary between the SM and  AFM phases is in agreement with
the mean field results~\cite{Hassan2013,Liang2014,Faye2014} for $t'<$0.8.
However it shows a distinguishable difference for $t'$ between 0.8 and 1.
Mean field calculations suggest that there is a QSL in this region~\cite{Hassan2013,Liang2014,Faye2014}, which is absent
in more rigorous fPEPS calculations.

 Figure \ref{fig:SpinOrder}(b) depicts the AFM and the zigzag order parameters along the line of $U=5$.
The system has an AFM order when $t'<$1, and a zigzag order when $t'>$1.15.
Both orders disappear  in between, which suggests that it is possibly a QSL phase.

\begin{figure} [tbp]
		\begin{center}
		\includegraphics[width=0.45\textwidth]{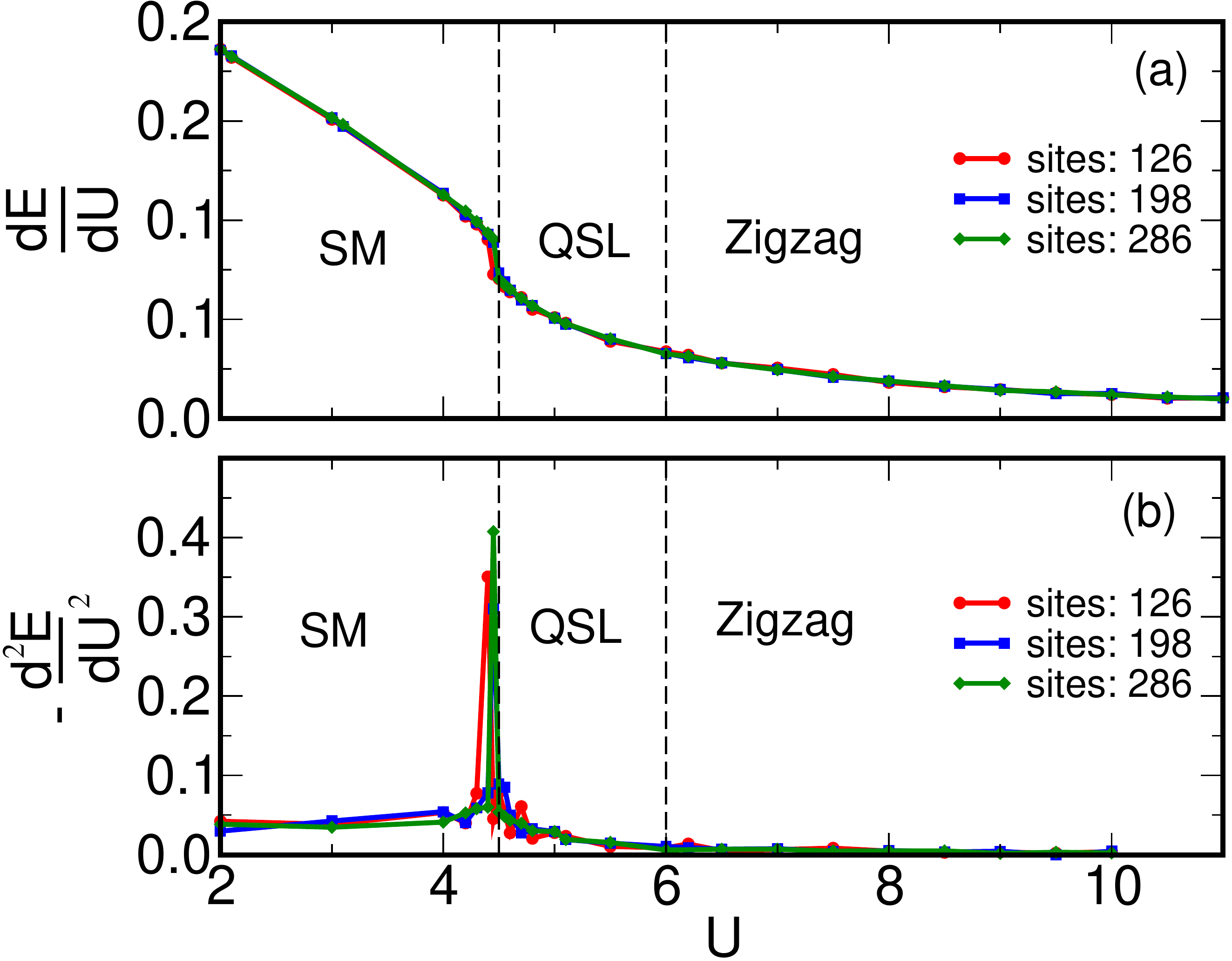}
		\caption{ (Color online) (a) The first-order energy derivative with respect to $U$ ($\frac{d E}{d U}$), and
(b) the second-order energy derivative ($\frac{d^2 E}{d U^2}$) on the lattices of different sizes with $t'$=1.05.
}
\label{fig:double_occupy}
		\end{center}
\end{figure}

To determine the order of the phase transitions, we calculate the first-order energy derivative with respect to $U$,
$\frac{d E}{d U}$, using the Hellmann-Feynman theorem, i.e.,
\begin{equation}
\frac{d E}{d U}=\langle \Psi |\frac{1}{N} \sum_i {\partial H \over \partial U} |\Psi \rangle=
\langle \Psi | \frac{1}{N}\sum_i  \hat{n}_{i,\uparrow}\hat{n}_{i,\downarrow}|\Psi \rangle,
\end{equation}
where $|\Psi\rangle$ is the ground state fPEPS wave function and $N$ is the total number of lattice sites that are used to calculate the total energy.
To reduce the boundary effects, only the central region of lattice is used to calculate the total energies
\cite{Stoudenmire2012,Liu2017,Liu2018}.
The results are shown in Fig.~\ref{fig:double_occupy}(a) for $t'$=1.05, and the second derivative of the energy with respect to $U$ (by the finite difference method)
are shown in Fig.~\ref{fig:double_occupy}(b). Clearly, there is a sharp discontinuity of $\frac{d E}{d U}$ at the SM-QSL transition around $U$=4.5, which suggests that
this is a strong first-order transition. In contrast, the transition between the QSL and zigzag phases is continuous.
Note that, $\langle \hat{n}_{i,\uparrow}\hat{n}_{i,\downarrow}\rangle$ also characterizes  the electron double occupancy on site $i$.
The QSL-SM transition is driven by the sudden increase of the double occupancy. The QSL phase also benefits from the increase in electron  double occupancy.

The values of the Heisenberg coupling $J$ and the Kitaev coupling $K$ in real materials, such as Na$_2$IrO$_3$ and $\alpha$-RuCl$_3$, have been estimated in Ref.\onlinecite{Sears2015,Banerjee2016,Shubhajyoti2019,Rau2014}. It has been suggested that in these materials, $|J|<|K|$ and $J/K<0$, i.e., $t'>t$.
We find a large region of zigzag phase in $t'>t$, $U>4.5$, which is consistent with the zigzag phase
in Na$_2$IrO$_3$ and $\alpha$-RuCl$_3$, determined by the resonant x-ray magnetic scattering and inelastic neutron scattering experiments\cite{Liu2011,Ye2012,Choi2012,Sears2015}. However, since the Iridium oxides are in the $t'>t$ region, it is a promising routine to find the QSL
in Iridium oxides that are close to the Mott insulator transitions.


To summarize, we calculate the ground state phase diagram of the Kitaev-Hubbard model at half-filling using the recently developed,
highly accurate fPEPS method. We obtain
the SM phase, AFM phase, and zigzag phase. Remarkably, we find a QSL phase
near the Mott insulator transition in the strong bond-dependent hopping region.
Our calculations suggest that it is possible to find the QSL in Iridium oxides, which have strong spin-dependent hopping
and are close to the Mott insulator transitions.


This work was supported by the National Natural Science Foundation of China (Nos. 12104433,11874343) and the Innovation program for Quantum Science and Technology (No. 2021ZD0301200).


%

\end{document}